ORIGINAL ARTICLE

# Towards emotion recognition for virtual environments: an evaluation of eeg features on benchmark dataset

M. L. R. Menezes[1]  · A. Samara[2] · L. Galway[2] · A. Sant'Anna[1] · A. Verikas[1] · F. Alonso-Fernandez[1] · H. Wang[2] · R. Bond[2]

**Abstract** One of the challenges in virtual environments is the difficulty users have in interacting with these increasingly complex systems. Ultimately, endowing machines with the ability to perceive users emotions will enable a more intuitive and reliable interaction. Consequently, using the electroencephalogram as a bio-signal sensor, the affective state of a user can be modelled and subsequently utilised in order to achieve a system that can recognise and react to the user's emotions. This paper investigates features extracted from electroencephalogram signals for the purpose of affective state modelling based on Russell's Circumplex Model. Investigations are presented that aim to provide the foundation for future work in modelling user affect to enhance interaction experience in virtual environments. The DEAP dataset was used within this work, along with a Support Vector Machine and Random Forest, which yielded reasonable classification accuracies for *Valence* and *Arousal* using feature vectors based on statistical measurements and band power from the $\alpha$, $\beta$, $\delta$, and $\theta$ waves and High Order Crossing of the EEG signal.

**Keywords** Affective computing · Virtual environment · EEG · Emotion recognition · Feature extraction

This work was funded by the Science Without Borders program from the Brazilian government and EU COST Action TD1405.

✉ M. L. R. Menezes
   maria.menezes@hh.se

1 Center for Applied Intelligent Systems Research,
  Halmstad University, Halmstad, Sweden

2 School of Computing and Mathematics,
  Ulster University Belfast, Belfast, UK

## 1 Introduction

Due to their increasing complexity, one of the main challenges found in virtual environments (VEs) is user interaction. Therefore, it is important to structure interaction modalities based on the requirements of the application, which may include both traditional and natural user interfaces, situational awareness and adaptation, personalised content management, multimodal dialogue and multimedia applications.

VEs typically require personalised interaction in order to maintain user engagement with the underlying task. While task engagement encompasses both the user's cognitive activity and motivation, it also requires an understanding of affective change in the user. Accordingly, physiological computing systems may be utilised to provide insight into the cognitive and affective processes associated with task engagement [15]. In particular, an indication of the levels of brain activity, through acquisition and processing of electroencephalogram (EEG) signals, may yield benefits when incorporated as an additional input modality [48].

In recent studies, EEG has been used to map the responses to the environment directly to the user's brain activity [1, 28, 33, 43, 49]. These systems are typically used for control purposes, enhancing traditional modalities such as mouse, keyboard, or game controller. However, this form of active interaction is still quite costly for users as it requires training and a good amount of both concentration and effort to modulate one's brain activity. This ultimately causes the user to focus more on the interaction modality itself than the underlying task. In order to achieve truly transparent interaction, the system is required to acquiesce to the user's intentions or needs. Using EEG as a bio-signal sensor to model the user's cognitive and affective state is one potential way to achieve an interaction that does not require any training or attention focus from the user.

Many authors have investigated the use of EEG for recognizing user affect. However, EEG signals are complex, multi-modal time series and there is no consensus on which features are better suited for this task. The main contributions of this paper are twofold: (1) a summary of how affect recognition can augment VR environments targeting different applications, namely, medicine, education, entertainment and lifestyle; (2) an evaluation of several types of features for affect recognition using EEG on a benchmark dataset. For the purposes of the investigations, the DEAP dataset was used to provide an annotated set of EEG signals [24]. Support Vector Machine (SVM) and Random Forest were employed to classify different affective states according to the Circumplex Model.

## 2 Background

A system that can detect and adapt to user's current affective state is interesting for a broad range of applications, from medicine and education to entertainment, games and lifestyle.

### 2.1 Applications in medicine

VEs have been shown to help in the treatment of many conditions, as well as help people cope with distressing emotions such as anxiety and stress. Virtual Reality Exposure Therapy (VRET), for example, is an increasingly common treatment for anxiety and specific phobias [36]. When a user is immersed in a VE, they can be systematically exposed to specific feared stimuli within a contextually relevant setting [4, 6, 16, 17]. VEs have also been shown to help children with Autism Spectrum Disorders (ASD) improve their social functioning [3]. These examples indicate where a system that uses emotional modulation could be useful: to help the physician analyse the emotional states and development of the patient's condition, as well as to use that information to adapt the treatment in real-time, avoiding possible over exposure of the patient.

### 2.2 Applications in education

The association between Affective Computing and learning is known as Affective Learning (AL): technologies that sense and respond to affective states during the learning process to make knowledge transfer and development more effective[41]. The recognition that interest and active participation are important factors in the learning process are largely based on intuition and generalization of constructivist theories [7, 41]. AL can change this scenario by measuring, modelling, studying and supporting the affective dimension of learning in ways that were not previously possible. Previous works have shown that VEs and AL can improve student performance [19, 27]. However, many of the previous approaches rely on questionnaires and other forms of off-line evaluation of affective state. The use of bio-sensors such as EEG might enable educational systems to automatic recognise affect and better understand non-verbal clues just as a teacher would.

### 2.3 Applications in entertainment and lifestyle

The entertainment industry is very enthusiastic regarding VEs, games being perhaps the most noticeable application. This enthusiasm is not surprising, to some degree, emotional experiences are what game designers create and sell [35]. Not only can VEs be designed to elicit both positive and negative emotions [13, 42], but also previous works have shown that emotion positively correlates with presence—the psychological sense of being in or existing in the VE in which one is immersed [2]. Another well-known use of VE in games are virtual worlds, such as Second Life [26]. The High Fidelity platform is able to track facial expressions in real time and transfer those to the user's avatar. Despite being able to mimic facial expressions related to speech and emotions, the system itself does not attempt to recognize affect [34]. EEG could extend the high fidelity platform with the ability to adapt to users' affect. It would also enable users that are unable to change their facial expressions—due to paralysis for example—to take advantage of a platform like High Fidelity. The use of the automatic modulation of user's emotional states in VEs are limitless and benefit from the proven relation between presence and emotional state.

### 2.4 EEG as an input modality for emotion recognition

Currently, various input modalities exist that can be utilised to acquire information about users and their emotions. More commonly, audiovisual-based communication, such as eye gaze tracking, facial expressions, body movement detection, and speech and auditory analysis may be employed as input modalities. Furthermore, physiological measurements using sensor-based input signals, such as EEG, galvanic skin response, and electrocardiogram can also be utilised. However, the use of EEG as an input modality has a number of advantages that make it potentially suitable for use in real-life tasks including its non-invasive nature and relative tolerance to movement. EEG can be used as a standalone modality as well as combined to other biometric sensors. The company *iMotions* for example has successfully developed a commercial platform for monitoring physiological and psychological parameters of users while experiencing VR. This is a great example of how affect recognition can be used to add value to VR applications [18, 20].

Several existing studies have exploited EEG as an input modality for the purpose of emotion recognition. Picard et al. looked at different techniques for feature extraction and selection in order to enhance emotion recognition by employing different biosignal data [40]. They found that there is a variation in physiological signals of the same subject expressing the same emotion from day to day. Which impairs recognition accuracy if not managed properly. Konstantinidis et al. studied real-time classification of emotions by analysing EEG data recorded using 19 channels. They showed that extracting features from EEG data using a complex non-linear computation, which is a multi-channel correlation dimension, and processing the features using a parallel computing platform (i.e. CUDA) would substantially reduce the processing time needed. Their method facilitates real-time emotion recognition [25].

Petrantonakis et al. proposes feature extraction methods based on Higher Order Crossing (HOC) analysis to recognise emotions from EEG data additionally to four different classification techniques. The highest reported classification accuracy was 83.33% using SVM trained on extracted HOC feature [37]. Murugappan investigated feature extraction using wavelet transforms [30]. Moreover, they used K-Nearest Neighbor to evaluate classification accuracy for emotions across two different sets of EEG channels (24 and 64 channels), with a resulting classification accuracy of 82.87%. Jenke et al. looked for feature selection methods extracted from EEG for emotion recognition [21]. They presented a systematic comparison of the wide range of available feature extraction methods using machine learning techniques for feature selection. Multivariate feature selection techniques performed slightly better than univariate methods, generally requiring less than 100 features on average.

Still there are challenges encountered when attempting to exploit EEG for emotional state recognition. Extracting relevant and informative features from EEG signals from a large number of subjects and formulating a suitable representation of this data in order to distinguish different affective states is an extremely complicated process [45]. This work utilizes a fairly large dataset of EEG signals to investigate the relevance of different features for the dimensions of Valence and Arousal, according to Russel's Circumplex Model of Affection. In this context, we aim to provide foundations for modelling user affect in order to enhance interaction experience in VEs.

# 3 Methodology

## 3.1 The DEAP dataset

The DEAP dataset [24], utilised in the work presented herein, comprises EEG and peripheral physiological signals for 32 subjects who individually watched 40 one-minute music videos of different genres as a stimulus to induce different affective states. Within the dataset 32 channels were used to record EEG signals for each trial per subject, resulting in 8064 samples that represent the signal over each one-minute trial. During each trial, a single subject rated his/her feelings after watching the video using the Self Assessment Manikin (SAM) scale in the range [1–9] to indicate the associated levels of Valence, Arousal, *Dominance* and *Liking*

The DEAP is a benchmark dataset for emotion analysis using EEG, physiological and video signals developed by researcher of the Queen Mary University of London, United Kingdom; the University of Twente, The Netherlands; the University of Geneva, Switzerland; and the École polytechnique Fédérale de Lausanne, Switzerland. Even though it does not represent data used in VEs per se, its data is considered consistent by more than 560 citations from the research community and a good source for affective data in general.

## 3.2 Selection of EEG channels

Psycho-physiological research has shown that left and right frontal lobes have significant activity during the experience of emotions [32]. There is also evidence of the role of the prefrontal cortex in affective reactions and particularly in emotion regulation and conscious experiment [12]. Many scientific experiments have successfully used electrodes located in those regions to analyse affective states [10, 37].

Since the purpose of this work is to model user affect aiming real time applications, a simpler and more user-friendly environment for data acquisition is required. In an effort to reduce the number of electrodes, the signals were selected from four positions Fp1, Fp2, F3 and F4 only, according to the 10–20 system, as seen in Fig. 1.

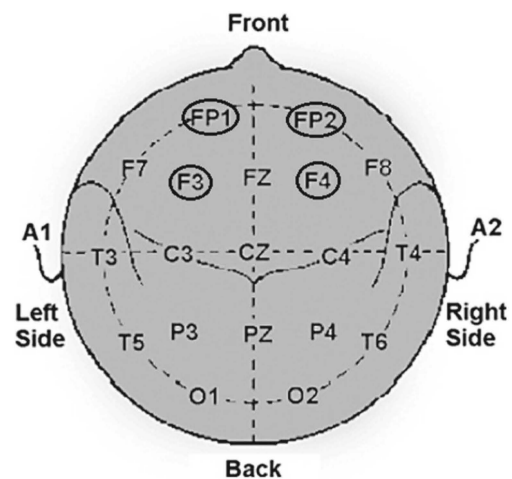

**Fig. 1** Fp1, Fp2, F3 and F4 positions selected according to the 10–20 system [31]

### 3.3 Bandwave extraction

Commonly, brainwaves are categorized into four different frequency bands: Delta ($\delta$) from 0.5 to 4 Hz; Theta ($\theta$) from 4 to 8 Hz; Alpha ($\alpha$) from 8 to 12 Hz; and Beta ($\beta$) 12 to 30 Hz. Literature has shown a strong correlation between these waves and different affective states [29].

The EEG data associated with each of the selected channels was transformed into $\alpha$, $\beta$, $\delta$, and $\theta$ waves, using the Parks–McClellan algorithm and Chebyshev Finite Impulse Response filter was applied to the signal according to the frequency ranges of each brainwave.

### 3.4 Feature extraction

Three types of features were computed from the EEG signal: statistical, powerband and High Order Crossing (HOC). Features along with the construction of the relevant feature vectors (FVS) are further explained within the following.

#### 3.4.1 Statistical features

We adopted six descriptive statistics, as suggested by Picard et al. in [40] and Petrantonakis in [38]. The statistical features were extracted from the EEG signal in time domain and from each of the brainwaves, creating separated feature vector for both time and frequency domain:

(a) Mean ($\mu$)
(b) Standard deviation ($\sigma$)
(c) Mean of the absolute values of the first differences ($AFD$)
(d) Mean of the normalised absolute values of the first differences ($\overline{AFD}$)
(e) Mean of the absolute values of the second differences ($ASD$)
(f) Mean of the normalised absolute values of the second differences ($\overline{ASD}$)

#### 3.4.2 Spectral power density of brain waves

For the selected four channels, the mean log-transformed brain wave power were extracted from the $\alpha$, $\beta$, $\delta$, and $\theta$ frequency bands, according to [11]. The Spectral Power Density (SPD) is widely used to detect the activity level in each brain wave, allowing the components in the frequency domain to be interpreted as electroencephalographic rhythms.

For each electrode was calculated four features, representing the medium power of the signal for the entire bandwave, result in a 16-feature vector:

$$FV_{SPD} = [f_{F_{p1}}, f_{F_{p2}}, f_{F_3}, f_{F_4}]$$

Being each channel feature ($f_{F_{ch}}$) a feature vector of the mean power of the signal for the respective bandwave:

$$f_{ch} = [SPD_\alpha, SPD_\beta, SPD_\delta, SPD_\theta]$$

#### 3.4.3 Higher order crossing

In this technique, a finite zero-mean time series $\{Z_t\}$, $t = 1, ..., N$ oscillating through level zero can be expressed by the number of zero crossings (NZC). Applying a filter to the time series generally changes its oscillation and consequently its NZC. When a specific sequence of filters is applied to a time series, a specific corresponding sequence of NZC is obtained. This is called a High Order Crossing (HOC) sequence [22, 38].

The difference operator ($\nabla$) is a high-pass filter defined as $\nabla Z_t \equiv Z_t - Z_{t-1}$. A sequence of filters $\Im_k \equiv \nabla_{k-1}$, $k = 1, 2, 3, ...$; and its corresponding HOC sequence, can then be defined as

$$D_k = NZC\{\Im_k(Z_t)\}, k = 1, 2, 3, ...; t = 1, ..., N$$

with

$$\Im(Z_t) = \nabla^{k-1} Z_t = \sum_{j=1}^{k} \frac{(k-1)!}{(j-1)!\,(k-1)!} (-1)^{j-1} Z_{t-j+1}$$

To calculate the number of zero-crossings, a binary time series is initially constructed given by:

$$X_t(k) = \begin{cases} 1, & \Im_k(Z_t) \geq 0 \\ 0, & \Im_k(Z_t) < 0 \end{cases}, k = 1, 2, 3, ...; t = 1, ...N$$

Finally, the HOC sequence is estimated by counting the symbol changes in $X_1(k), ..., X_N(k)$:

$$D_k = \sum_{t=2}^{N} [X_t(k) - X_{t-1}(k)]^2$$

In this paper filters up to order six were used, creating the feature vector $FV_{HOC} = [D_1, D_2, ..., D_6]$.

### 3.5 Affective state classification

The Circumplex Model of Affect developed by James Russell suggests that the core of emotional states are distributed in a two-dimensional circular space, containing Arousal and Valence dimensions. Arousal is represented by the vertical axis and Valence is represented by the horizontal axis, while the center of the circle represents a neutral level of Valence and Arousal [44], as seen in Fig. 2.

As the current study is interested in recognising the affective state that a subject is experiencing, congruous with the two-dimensional Russell's Circumplex Model, throughout the investigations only Valence and Arousal ratings were used. Valence and Arousal ratings are provided within the DEAP dataset as numeric values ranging from [1–9] based on the SAM scale [5]. Two different partitioning schemes have been employed in order to discretize the range of values within the scale, as illustrated in Fig. 3, and given as follows:

(a) *Tripartition Labeling Scheme*: Dividing the scale into three ranges [1.0–3.0], [4.0–6.0] and [7.0–9.0], given as the partitions *Low, Medium* and *High* respectively.

(b) *Bipartition Labeling Scheme*: Similar to the previous scheme, however we removed instances annotated as *Medium*, resulting in the two ranges [1.0–3.0] and [7.0–9.0], given as the partitions *Low* and *High* respectively.

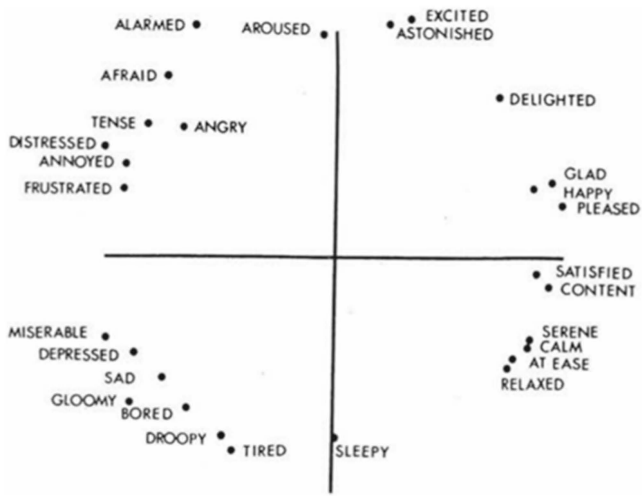

**Fig. 2** Russel's Circumplex Model of Affect [44]

Within the research literature, a range of classification techniques have been used for affective computing and emotion recognition using EEG bio-signals as an input modality [23]. For the investigations presented herein we utilised two different classification methods: C-Support Vector Classification (SVM) with a linear kernel and Random Forest. The chosen SVM implementation is available from the LIBSVM library developed at National Taiwan University [9, 14] and the Random Forest developed by Leo Breiman [8].

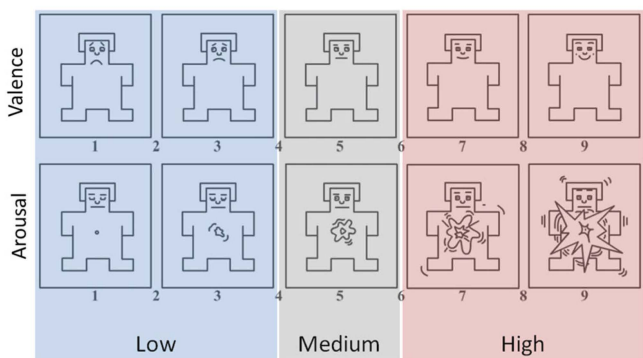

**Fig. 3** Mapping from SAM scale Valence and Arousal values to Labels (*Low, Medium, High*)

Support Vector Machine (SVM) and Random Forest (RF) are versatile and widely used methods that have been shown to perform well in many application areas. The success of SVMs have been attributed to three main reasons: "their ability to learn well with only a small number of free parameters; their robustness against several types of model violations and outliers; and their computational efficiency compared to other methods." [46]. Compared to other machine learning methods, RF present three interesting additional features: "a built-in performance assessment; a measure of relative importance of descriptors; and a measure of compound similarity that is weighted by the relative importance of descriptors" [47].

## 4 Experimental results

For the sake of exploration of different features, as previously described, we used classification accuracy as a metric. Furthermore, we have utilised the 10-fold cross validation approach for assessing classification performance. As previously discussed, this investigation aims to identify patterns related to features extracted from EEG signals across different Valence and Arousal states. For that, we applied SVM and Random Forest. Moreover, two labeling schemes were employed for each of the affective dimensions, i.e. *Bipartition* and *Tripartition*.

The following tables show the average results obtained for all the instances in the dataset, i.e. all videos for all participants. A comparison of the SVM and Random Forest results for all methods can be seen in Tables 1 and 2.

We can see that the results obtained for Random Forest were slightly better than SVM for all methods except Spectral Power Density. The comparison of the two tables show that the features extracted from the EEG signal behave similarly for any of the classification methods applied. Being the biggest difference for Statistic features extracted from Brainwaves for Arousal, that for Random Forest had 74.0% accuracy and for SVM only 57.2%. We can conclude that Random Forest performed better for all the Features in general and specially for Statistics of Brainwaves. SVM can be a better choice if the chosen features are the spectral power density of Brainwaves and the class in interest is Valence.

**Table 1** Classificationt accuracy per method, using SVM

| Method | Bipartition | | Tripartition | |
|---|---|---|---|---|
| | Arousal | Valence | Arousal | Valence |
| Statistics—Time | 65.0% | 61.2% | 57.0% | 51.3% |
| Statistics—Bandwaves | 57.2% | 83.2% | 59.7% | 55.1% |
| Bandwaves SPD | 69.2% | 88.4% | 59.5% | 55.9% |
| HOC | 56.8% | 62.7% | 59.1% | 53.5% |

**Table 2** Classification accuracy per method, using Random Forest

| Method | Bipartition | | Tripartition | |
|---|---|---|---|---|
| | Arousal | Valence | Arousal | Valence |
| Statistics—Time | 67.1% | 61.3% | 57.7% | 50.0% |
| Statistics—Bandwaves | 74.0% | 88.4% | 63.1% | 58.8% |
| Bandwaves SPD | 67.9% | 86.6% | 56.1% | 55.2% |
| HOC | 57.4% | 64.7% | 57.8% | 55.1% |

**Table 4** Classification accuracy for all Statistical Features for each Brainwave using Random Forest

| Statistics—Bandwaves | Bipartition | | Tripartition | |
|---|---|---|---|---|
| | Arousal | Valence | Arousal | Valence |
| $\alpha$ | 66.6% | 78.3% | 59.7% | 54.0% |
| $\beta$ | 67.5% | 77.1% | 59.3% | 54.2% |
| $\delta$ | 75.7% | 87.9% | 60.9% | 55.3% |
| $\theta$ | 71.3% | 84.9% | 61.3% | 53.5% |

Tables 1 and 2 show that Bipartition overcomes Tripartition for all methods tested except Arousal for HOC. Although the approximately 2% for Bi and Tripartition do not represent a statistically significant difference in accuracy. The best result for Tripartition is 63.1% for the Statistic features of the Brainwaves and Arousal in Table 2. Despite the results for Arousal in Tripartition being slightly better than the ones for Valence, the difference is not statistically significant.

The results are more interesting for Bipartition, in which the features tested are generally better representatives for Valence than Arousal, with an average difference of approximately 9% and the highest difference of approximately 18% for SPD.

We can also note that the best results were obtain for the methods that involve Bandwaves' features: Statistics and SPD. Valence has the best accuracies of 88.4 and 86.6%, respectively. The result for Arousal are 74.0 and 67.9% in Table 2.

**Table 3** Classification accuracy for SPD using SVM

| SPD | Bipartition | | Tripartition | |
|---|---|---|---|---|
| | Arousal | Valence | Arousal | Valence |
| $\alpha$ | 52.6% | 73.1% | 59.0% | 57.3% |
| $\beta$ | 64.6% | 69.8% | 59.0% | 55.6% |
| $\delta$ | 66.2% | 82.9% | 60.2% | 54.9% |
| $\theta$ | 62.9% | 76.1% | 59.4% | 55.9% |
| $\alpha, \beta$ | 65.6% | 82.7% | 59.8% | 56.1% |
| $\alpha, \delta$ | 66.5% | 88.1% | 59.4% | 57.7% |
| $\alpha, \theta$ | 62.1% | 83.4% | 58.8% | 55.1% |
| $\beta, \delta$ | 66.9% | 88.4% | 59.1% | 56.3% |
| $\beta, \theta$ | 67.5% | 85.4% | 59.7% | 55.6% |
| $\delta, \theta$ | 67.1% | 88.9% | 59.8% | 55.2% |
| $\alpha, \beta, \delta$ | 67.5% | 88.6% | 59.9% | 57.2% |
| $\alpha, \beta, \theta$ | 65.8% | 85.0% | 59.5% | 56.8% |
| $\alpha, \delta, \theta$ | 66.2% | 88.7% | 59.4% | 57.0% |
| $\beta, \delta, \theta$ | 67.7% | 88.4% | 59.2% | 54.2% |
| $\alpha, \beta, \delta, \theta$ | 69.2% | 88.4% | 59.5% | 55.9% |

The subsequent tables show the percentage of correctly classified instances for the methods that showed the best results: SPD using SVM, in Table 3; and Statistics of Brainwave using Random Forest, in Tables 4 and 5.

In Table 3 is clear that SPD features best relate to Valence in Bipartition, being $\delta$'s SPD the best single feature with 82.9% accuracy. Combining two other features, such as the SPD of $\alpha$ and $\beta$ or $\alpha$ and $\theta$ or even $\beta$ and $\theta$ we can obtain similar results as $\delta$ alone: 82.7, 83.4 and 85.4%, respectively. Combining any of the single features with $\delta$'s SPD increases the accuracy approximately 5%. The second best single feature is $\theta$'s SPD and combining both $\delta$'s and $\theta$'s SPD gives the best result of 88.9%, better than combining all features in one single vector, 88.4%.

Table 4 shows the accuracy obtained for the Statistic features of each of the single Brainwaves using Random Forest. Being $\delta$ and $\theta$ again the brainwaves which features have the best results, 87.9 and 84.9% for Valence and 75.7 and 71.3% for Arousal in Bipartition. Combining the statistical features of the two bandwaves $\delta$ and $\theta$ increases the accuracy for Valence to 88.2%, almost the same as using the features for all bandwaves, 88.4%. Combining those same features for Arousal, on the other hand, gives the accuracy of 73.8%, worse than the result for $\delta$ only.

Table 5 shows the accuracy obtained for each of the single Statistic features for all Brainwaves combined using Random Forest. Here, we can see again the best results for Valence in Bipartition. For the single statistical features,

**Table 5** Classification accuracy for all Brainwaves for each Statistical feature, using Random Forest

| Statistics—Bandwaves | Bipartition | | Tripartition | |
|---|---|---|---|---|
| | Arousal | Valence | Arousal | Valence |
| $\mu$ | 64.1% | 80.7% | 55.9% | 53.5% |
| $\sigma$ | 70.7% | 87.4% | 56.9% | 54.8% |
| $AFD$ | 64.8% | 89.9% | 54.8% | 54.9% |
| $\overline{AFD}$ | 68.3% | 71.3% | 58.4% | 53.5% |
| $ASD$ | 67.7% | 88.4% | 57.7% | 55.5% |
| $\overline{ASD}$ | 71.7% | 73.8% | 57.9% | 53.8% |

*AFD* has the best result of 89.9%, followed by *ASD* and $\sigma$, with 88.4 and 87.4%, respectively. Combining the three set of features again does not give a better accuracy than the best single feature, resulting in 88.6%.

For Valence, on the other hand, the best features are $\overline{ASD}$, $\sigma$, $\overline{AFD}$ and *ASD*, with 71.7, 70.7, 68.3 and 67.7% classification accuracy, respectively. Combining those features does not improve the accuracy, resulting in 68.8% classification accuracy.

# 5 Discussion

The investigations and associated results presented in this paper show the potential of utilizing EEG signal data for recognising affective states. Based on the classification accuracy, the approach could be used to effectively recognise emotions in certain types of virtual reality environments. Educational applications could benefit from it by adapting the content of a course to the students anxiety levels, characterised by low levels of arousal and valence, detected by the bipartition approach. Other than that, the approach presented could be applied to medical applications that aim to help patients deal with phobias or entertainment platforms for social anxiety.

Both classification methods applied gave similar results, being the results for Random Forest slightly better than the ones for SVM. Particularly, the highest classification accuracy was obtained using the feature vector generated based on the statistical measurements derived from brainwaves, e.g 88.4% for Valence and 74% for Arousal.

Likewise, using a feature vector based on the associated power bands and SVM also produced the classification accuracy of 88.4% for Valence and slightly lower for Arousal, 69.2%. In both cases, the *Bipartition* labelling scheme was used.

For both methods of feature extraction, the features associated with $\delta$ and $\theta$ performed better than the other bandwaves. The best accuracy obtained was for the combination of the SPD for both $\delta$ and $\theta$, resulting in 88.9% correctly classified instances.

The features that can be better associated with the affective state of Valence are $\overline{ASD}$, $\sigma$, $\overline{AFD}$ and *ASD*, with 71.7, 70.7, 68.3 and 67.7%.Combining those features does not improve the results.

The highest classification accuracy rates were obtained using features extracted from the brainwaves, corroborating the neurophysilogical theories that relate those with several different mental states. The Statistic features and Spectral Power Density represent the activity level in each bandwave and can give us an insight about the relation between the affective dimensions of Valence and Arousal and the brain activation in each frequency. In Figs. 4 and 5, we can see the Receiver Operating Characteristic (ROC) curve for both Statistic features and SPD respectively.

The red dashed line represents the equivalent of a random guess. The higher the curves are from this diagonal, the more sensitive it is regarding the class, Valence or Arousal. Analysing those curves for Valence we can see in Fig. 4a that $\sigma$, *AFD* and *ASD* have the best results, as well as $\delta$ and $\theta$ in Fig. 5a, corroborating the results obtained from both classification methods. The results for Arousal show curves close to the diagonal, again corroborating our previous results, of low accuracy for all methods in general.

Figures 6 and 7 show the distribution of features providing worst accuracy, in Fig. 6 and best accuracy, in Fig. 7. We can see that the features of the methods with worst accuracy, such as the HOC features and the Statistical features in time domain, overlap for the classes of High and Low Valence and Arousal. On the contrary, there is less overlap on the features obtained from the methods with best accuracy, such as $\sigma$ and SPD of Brainwaves.

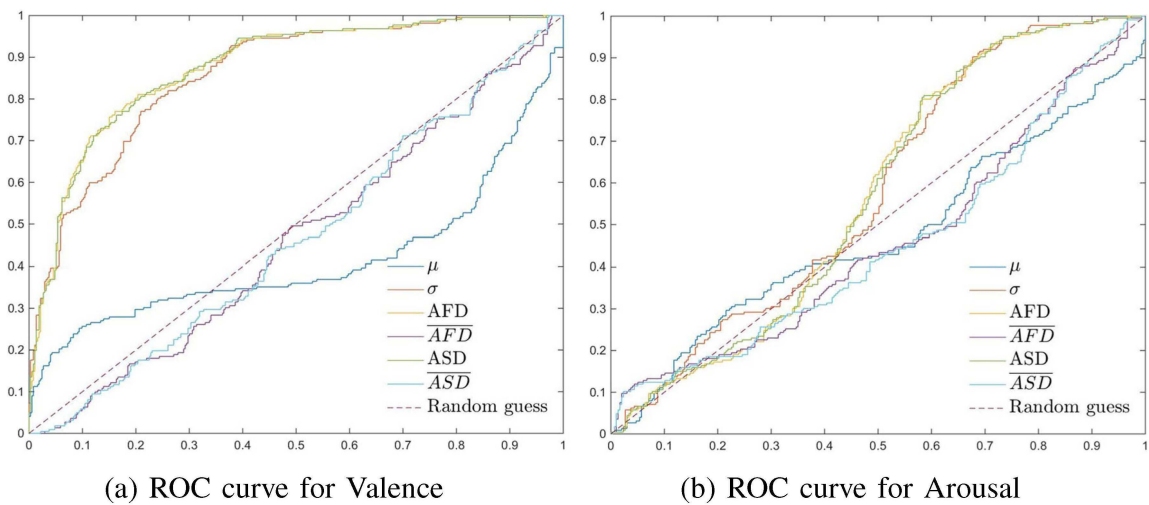

(a) ROC curve for Valence      (b) ROC curve for Arousal

**Fig. 4** ROC curves for each of the Statistical Features for Brainwaves

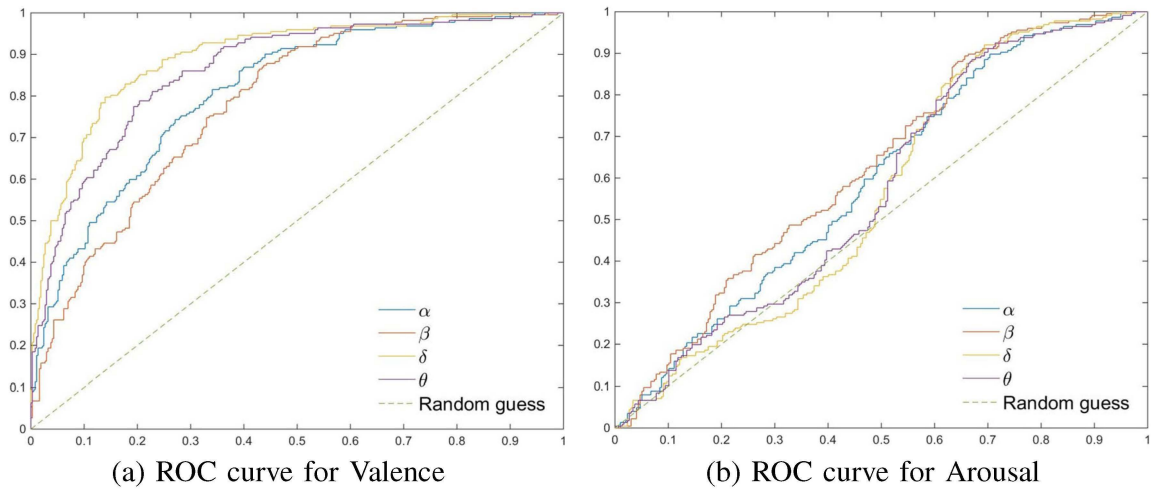

(a) ROC curve for Valence     (b) ROC curve for Arousal

**Fig. 5** ROC curves for the SPD of each of the Brainwaves.

Even though we obtained good results for some features, we can see in Figs. 6 and 7 that histograms of even the best features overlap considerably and result in ROC curves close to Random guess, as seen in Figs. 4 and 5. This characteristics observed amongst the features investigated could be due to many reasons. The sensitivity of the self-assessment scale used to garner affect ratings is subjective, as it is based on the thoughts and impressions of the participant about the video he/she watched. Moreover, it is often the case that people do not know how to articulate their actual emotions and associated states due to ambiguity and mixed mental activities [39].

Therefore, it is potentially the case that some of the participants could not precisely entail their actual emotional state using the SAM scale. Due to this factor, classification models were generated twice using two different mapping schemes in order to determine the impact from ambiguous annotations that potentially arise from the selection of Valence and Arousal values from the middle of the self-assessment scale. As the results indicated, placing such a constraint on the ranges of affect to be modelled improved the overall classification performance.

In the majority of the investigations, the classification accuracies obtained for Valence outperformed those obtained for Arousal. It is difficult to determine why this was the case but several factors may have contributed to this effect. One possible reason is that the concept of Arousal may be more difficult to understand and categorize than Valence, resulting in inconsistent labelling. In addition, participants within the DEAP dataset watched video clips as a stimuli, hence were passive during that time, resulting in a small range of Arousal values that were not distinctive

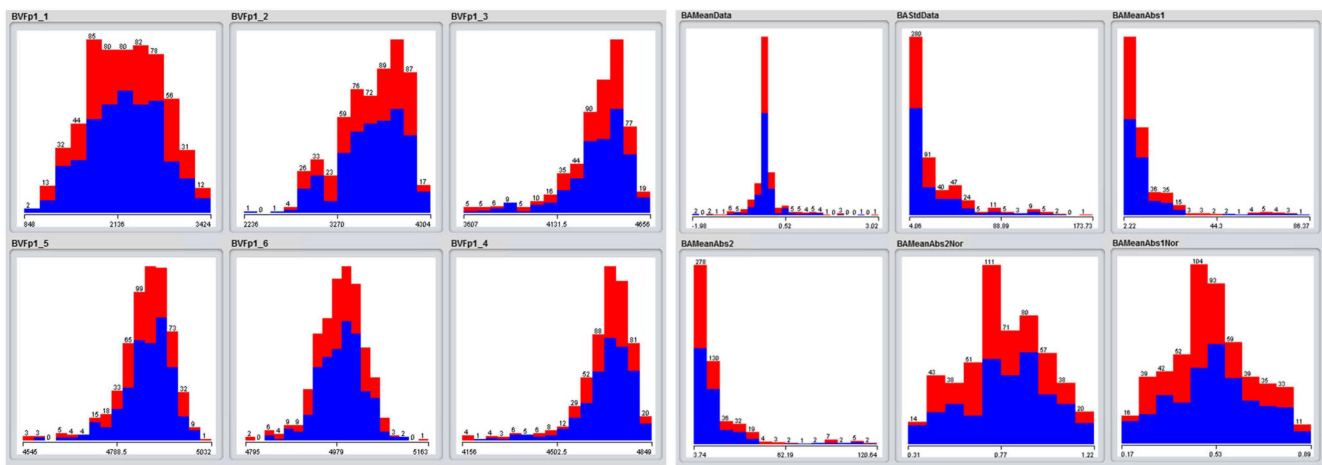

(a) Histogram of each of the 6 filter's HOC features for channel Fp1 and Valence    (b) Histogram of each of the 6 Statistical features in Time for Arousal

**Fig. 6** Histograms of the features providing worst accuracy using Bipartition: the High class in *red* and Low in *blue*

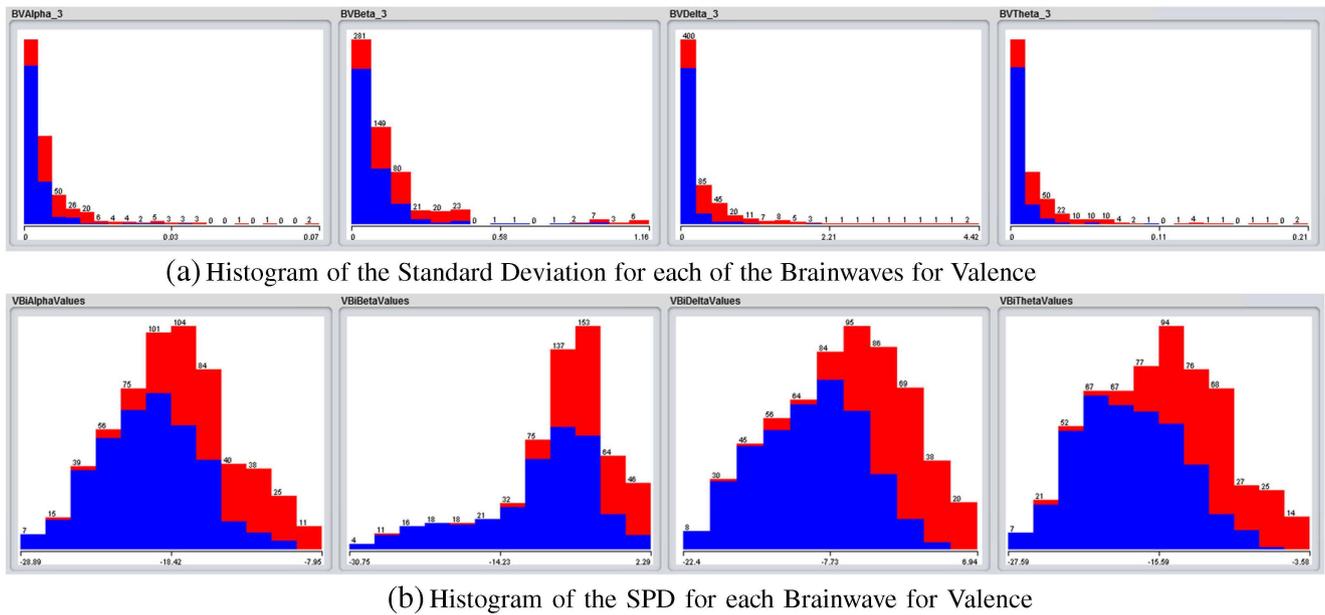

(a) Histogram of the Standard Deviation for each of the Brainwaves for Valence

(b) Histogram of the SPD for each Brainwave for Valence

**Fig. 7** Histograms of the features providing best accuracy using Bipartition: the High class in *red* and Low in *blue*

enough to be picked up by the classifier. This specific aspect could be improved if the data were obtained using a virtual environment, where the person has a greater sense of presence, hence having more influence in their emotional state, as discussed in chapter 2.

# 6 Conclusions and future work

This paper investigated exploiting electroencephalogram data as an input modality for the purpose of providing VEs with the ability to recognize and detect the emotional states of users. Consequently, the results from several experiments using different sets of features, especially the ones descendant from brainwaves, extracted from EEG data within the DEAP dataset show the potential of utilizing EEG signal data.

In addition, the observed discrepancy in classification accuracy due to different affective state mapping schemes was discussed, indicating that a degree of ambiguity will exist within such datasets, which has an obvious effect on the ability to accurately model affective states.

Moreover, combining several features together does not necessarily increase classification accuracy, as discussed in chapter 4. For example, as shown in Table 4, Combining $\delta$'s and $\theta$'s statistical features for Arousal, gives the accuracy of 73.8%, worst than the result for $\delta$ only.

Additionally, as the results depict, the features extracted from $\alpha$, $\beta$, $\delta$ and $\theta$ waves and the classification accuracies obtained for Valence makes it potentially suitable as a metric for measuring this aspect of the affective state of a user, ranging from negative to positive (i.e. *Low-Valence* to *High-Valence*).

The preliminary results shown in this article will help informing and leading to further experiments that eventually integrate different input modalities together with EEG in order to potentially provide a more robust model of the user's affective state. The current set of investigations would benefit if repeated using another mapping scheme based on Fuzzy Logic, for example, in an effort to improve the classification of potentially ambiguous affective states.

It is also interesting to extend the investigation regarding brain activation and the affective dimensions of Valence and Arousal. Not only how the negative (*Low-Valence*) and positive (*High-Valence*) states relate in terms of absolute values with the brain activation, but also how this activation is propagated though the entire extension of the brain.

Nonetheless, it is important to expand the study and the methods to real-time applications, and determine how those might behave in the real scenario of VEs. Not only taking into account the computational cost, aiming for real-time and embedded systems; but also how the virtual environment should adapt to this new form of awareness and how the user will react to this new form of enhanced interaction.

The article also discusses the importance of taking into account the effective qualities of the virtual environment to improve user experience and the many potential applications of such awareness for a different range of areas, such as medicine, education, entertainment and life style. The affective qualities of a virtual environment contribute to the engagement or feeling of presence of the user and vice-versa. When the affective qualities of the VE do not match the expectations of the user or the affective level of the situation being lived at the environment, it may have a negative effect on the user experience. Recognizing the importance of the affective qualities and awareness of a VE and introducing these often neglected aspects into the development process will improve the user experience.

**Acknowledgements** The authors would like to thank COST for supporting the work presented in this paper (COST-STSM-TD1405-33385) and CNPq for the Science Without Borders Scholarship.